\documentclass[twocolumn,english]{revtex4-1}
\usepackage{mathptmx}

\usepackage[T1]{fontenc}
\usepackage[latin9]{inputenc}
\usepackage{amsmath}
\usepackage{amssymb}
\usepackage{graphicx}

\makeatletter

\providecommand{\tabularnewline}{\\}

\makeatother

\usepackage{babel}
\begin{document}
\title{Spinteract: A Program to Refine Magnetic Interactions to Diffuse Scattering
Data}
\author{Joseph A. M. Paddison}
\email{paddisonja@ornl.gov}

\affiliation{Materials Science and Technology Division, Oak Ridge National Laboratory,
Oak Ridge, TN 37831, USA}
\affiliation{Churchill College, University of Cambridge, Storey's Way, Cambridge,
CB3 0DS, U.K.}
\begin{abstract}
Magnetic diffuse scattering---the broad magnetic scattering features
observed in neutron-diffraction data above a material's magnetic ordering
temperature---provides a rich source of information about the material's
magnetic Hamiltonian. However, this information has often remained
under-utilised due to a lack of available computer software that can
fit values of magnetic interaction parameters to such data. Here,
an open-source computer program, Spinteract, is presented, which enables
straightforward refinement of magnetic interaction parameters to powder
and single-crystal magnetic diffuse scattering data. The theory and
implementation of this approach are summarised. Examples are presented
of refinements to published experimental diffuse-scattering data sets
for the canonical antiferromagnet MnO and the highly-frustrated classical
spin liquid Gd$_{3}$Ga$_{5}$O$_{12}$. Guidelines for data collection
and refinement are outlined, and possible developments of the approach
are discussed.
\end{abstract}
\maketitle

\section{Introduction}

Magnetic materials show a wide variety of interesting and important
phenomena, ranging from spin-liquid phases to multiferroicity \citep{Broholm_2020,Balents_2010,Cheong_2007}.
A prerequisite for understanding such behaviour is often to determine
the underlying magnetic interactions. The ``gold standard'' approach
to achieve this aim typically involves performing inelastic neutron
scattering experiments on a material below its magnetic ordering temperature
$T_{N}$, and then fitting an interaction model to the observed spin-wave
(magnon) spectra using linear spin-wave theory \citep{Kubo_1952,Pepy_1974,Princep_2017,Fishman_2018}.
This approach was pioneered in the 1950s and 1960s, and has since
been successfully applied to a wide range of materials. The recent
development of powerful yet user-friendly computer software to perform
spin-wave calculations, such as the SpinW code \citep{Toth_2015},
has played a crucial role in popularising the application of the approach.

The spin-wave approach to determining magnetic interactions is highly
successful, but nevertheless has certain limitations. First, it is
often relatively time-consuming, because two different neutron-scattering
experiments are typically required---diffraction measurements to
determine the ordered magnetic structure of a material, followed by
inelastic neutron-scattering measurements to parametrise its magnetic
interactions. Second, the application of LSWT assumes a magnetic state
with long-range order \citep{Kubo_1952,Fishman_2018}. It is therefore
unsuitable for materials that do not show long-range magnetic order
at experimentally-accessible temperatures, such as those with spin-liquid
or spin-glass ground states \citep{Broholm_2020,Balents_2010,Zhou_2017}.
It also requires that the magnetic structure is well understood, which
may not be the case for complex phases such as non-coplanar spin textures.
Third, robust fitting of magnetic interactions to measured magnon
intensities is computationally expensive; this is true in particular
for data collected on powder rather than single crystal samples, due
to the requirement to spherically average the calculated intensities. 

An alternative approach to determine magnetic interactions employs
neutron-diffraction measurements performed \emph{above} $T_{N}$.
In the paramagnetic phase, spin-pair correlations are short-ranged
in real space and therefore give rise to broad scattering features,
which are known as magnetic diffuse scattering. It has long been recognised
that diffuse-scattering data are sensitive to the underlying magnetic
interactions; e.g., in 1964, Blech and Averbach studied the magnetic
diffuse scattering of the prototypical antiferromagnetic MnO above
its $T_{N}$ \citep{Blech_1964}, and extracted an estimate of the
dominant exchange interaction in good agreement with recent results
\citep{Hohlwein_2003}. Since that time, magnetic diffuse-scattering
experiments have provided valuable insight into the magnetic interactions
in materials such as spin ices \citep{Bramwell_2001a,Fennell_2009,Yavorskii_2008,Sibille_2018},
spin-liquid candidates \citep{Paddison_2013a,Paddison_2017,Bai_2019,Li_2017a,Gao_2017,Chillal_2020},
and skyrmion crystals \citep{Gao_2020,Paddison_2022}. Furthermore,
it was recently shown that diffuse-scattering data are also sensitive
to bond-dependent magnetic interactions \citep{Paddison_2020}, such
as those implicated in the celebrated Kitaev model \citep{Kitaev_2003}.
In all cases, the reason for this sensitivity is that the magnetic
diffuse scattering changes continuously as the interaction space is
traversed. By contrast, the magnetic Bragg scattering measured below
$T_{N}$ provides little information about the underlying interactions,
because the same ordered magnetic structure is typically obtained
throughout wide regions of interaction space. 

Traditionally, diffuse-scattering analysis approaches were developed
specifically for the problem at hand, because general-purpose software
was not available. Recently, programs to analyse magnetic diffuse-scattering
data have been developed that fit local spin arrangements directly
to experimental data, using either ``big box'' methods such as reverse
Monte Carlo refinement \citep{Tucker_2007,Mellergard_1998,Paddison_2013,Harcombe_2016,McGuire_2016,Nilsen_2015,Paddison_2018,Morgan_2021},
or ``small box'' methods such as magnetic pair-distribution function
analysis \citep{Frandsen_2014,Frandsen_2015,Andersen_2021,Baral_2022,Frandsen_2022}.
These approaches aim to describe the correlations between spin pairs
using methods familiar from crystal-structure refinement, and have
provided important insights into the physics of disordered magnetic
states, from magnetic nanoparticles \citep{Andersen_2021} to emergent
partial magnetic ordering \citep{Paddison_2016}. However, such approaches
are limited in the sense that they aim to determine only the spin
correlations---not the magnetic interactions that drive them. 

Here, I present a computer program, Spinteract, which calculates magnetic
diffuse scattering data from interaction models and fits interaction
parameter values directly to experimental data. The Spinteract program
can analyse multiple data sets simultaneously, including powder diffuse
scattering, single-crystal diffuse scattering, and bulk magnetic susceptibility.
This article is structured as follows. I begin by introducing the
types of spin Hamiltonian that can be modeled using Spinteract, and
the theory used to calculate the magnetic diffuse scattering from
the interaction parameters. I summarise the refinement procedure in
Spinteract. I benchmark the program using previously-published experimental
data on two well-studied magnetic materials---the canonical antiferromagnet
MnO \citep{Blech_1964,Pepy_1974,Hohlwein_2003,Frandsen_2015,Paddison_2018}
and the classical spin liquid Gd$_{3}$Ga$_{5}$O$_{12}$ \citep{Kinney_1979,Schiffer_1995,Petrenko_1998,Yavorskii_2006,Paddison_2015,dAmbrumenil_2015}---and
show that the analysis is consistent with published results, and also
allows testing of models that extend the previous analyses. I then
summarise experimental considerations that enable accurate measurement
of diffuse scattering patterns, and successful refinement strategies.
I conclude by discussing potential applications and developments of
this approach.

\section{Theory\label{sec:Theory}}

\subsection{Spin Hamiltonian}

Let us first introduce the types of spin Hamiltonians that are considered
in Spinteract. Throughout, I consider a crystal with $N$ magnetic
atoms in its primitive unit cell; for simplicity, it is assumed that
these atoms are crystallographically equivalent. A spin in the crystal
is denoted $\mathbf{S}_{i}(\mathbf{r})$, where $i\in\left\{ 1,N\right\} $
labels an atomic position $\mathbf{R}_{i}$ in the primitive unit
cell, and $\mathbf{r}$ is a lattice vector giving the origin of this
unit cell within the crystal. The spin quantum number is $S$ and,
in the classical approximation, the spins are taken as vectors of
length $\sqrt{S(S+1)}$ in the spin-only case. The $g$-factor is
assumed to be isotropic, so that the magnetic dipole moment $\text{\textbf{m}}=g\mathbf{S}$,
but the extension to anisotropic $g$-factors is straightforward.
Throughout, I will consider only bilinear interactions. 

The spin Hamiltonian in zero applied magnetic field can be written
as the sum of a single-ion term and $H_{\mathrm{si}}$ and a pairwise
interaction term $H_{\mathrm{ex}}$, 
\begin{equation}
H=H_{\mathrm{si}}+H_{\mathrm{ex}}.
\end{equation}
In many magnetic materials, especially insulating compounds containing
magnetic transition-metal ions, pairwise interactions can be approximated
by the isotropic (Heisenberg) form,
\begin{equation}
H_{\mathrm{ex,iso}}=-\frac{1}{2}\sum_{i,\mathbf{r}}\sum_{j,\mathbf{r}^{\prime}}J_{ij}^{\mathrm{iso}}(\mathbf{r}^{\prime}-\mathbf{r})\mathbf{S}_{i}(\mathbf{r})\cdot\mathbf{S}_{j}(\mathbf{r}^{\prime}),\label{eq:heisenberg}
\end{equation}
where $J_{ij}^{\mathrm{iso}}(\mathbf{r}^{\prime}-\mathbf{r})$ is
the Heisenberg interaction parameter for the bond connecting site
$i$ in primitive cell $\mathbf{r}$ with site $j$ in primitive cell
$\mathbf{r}^{\prime}$. Usually, these interaction are restricted
to near-neighbour distances, such that $J_{ij}(\mathbf{r}^{\prime}-\mathbf{r})\in\left\{ J_{1},J_{2},...,J_{n}\right\} $,
where $J_{n}$ denotes an interaction between $n$-th nearest neighbours.
Spinteract uses the convention that ferromagnetic interactions correspond
to positive values of $J$, and each pair of spins is counted once
in the double summation.

The Heisenberg form can be generalised to include anisotropic interactions,
\begin{equation}
H_{\mathrm{ex}}=-\frac{1}{2}\sum_{i,\mathbf{r}}\sum_{j,\mathbf{r}^{\prime}}\sum_{\alpha,\beta}J_{ij}^{\alpha\beta}(\mathbf{r}^{\prime}-\mathbf{r})S_{i}^{\alpha}(\mathbf{r})S_{j}^{\beta}(\mathbf{r}^{\prime}),
\end{equation}
where $\alpha,\beta\in{x,y,z}$ denote spin components with respect
to a set of local principal axes (discussed below). This can include,
for example, the XXZ model ($J^{xx}=J^{yy}\neq J^{zz}$) as well as
interactions that depend on the orientations of bonds connecting spin
pairs, such as the Kitaev interaction \citep{Kitaev_2003}.

A special case is the long-ranged magnetic dipolar interaction, given
by
\begin{equation}
H_{\mathrm{dip}}=\frac{g^{2}D_{\mathrm{dip}}\left|\mathbf{r}_{\mathrm{nn}}\right|^{3}}{2}\sum_{i,\mathbf{r}}\sum_{j,\mathbf{r}^{\prime}}\frac{\left\{ \mathbf{S}_{i}(\mathbf{r})\cdot\mathbf{S}_{j}(\mathbf{r}^{\prime})-3[\mathbf{S}_{i}(\mathbf{r})\cdot\hat{\mathbf{r}}_{ij}][\mathbf{S}_{j}(\mathbf{r}^{\prime})\cdot\hat{\mathbf{r}}_{ij}]\right\} }{\left|\mathbf{r}_{ij}\right|^{3}},
\end{equation}
where\textbf{ $\mathbf{r}_{ij}$ }is a unit vector parallel to the
vector connecting spin pairs in the crystal, and $D_{\mathrm{dip}}=\mu_{0}\mu_{\mathrm{B}}^{2}/4\pi k_{\mathrm{B}}|\mathbf{r}_{\mathrm{nn}}|^{3}$
is the magnitude of the dipolar interaction at the nearest-neighbour
distance, $|\mathbf{r}_{\mathrm{nn}}|$. The magnetic dipolar interaction
is usually negligible compared to the exchange interaction in transition-metal
compounds, but can be significant in rare-earth compounds, where exchange
interactions are often weak but magnetic moments can have large magnitudes.
In Spinteract, the magnetic dipolar interaction is implemented using
Ewald summation, as described in Ref.~\citep{Enjalran_2004}.

The form of the single-ion Hamiltonian $H_{\mathrm{si}}$ depends
on the point symmetry of the magnetic site. For cubic site symmetries,
no bilinear single-ion term is allowed. For lower site symmetries,
the single-ion Hamiltonian is given by
\begin{equation}
H_{\mathrm{si}}=-D\sum_{i,\mathbf{r}}[S_{i}^{z}(\mathbf{r})]^{2}-E\sum_{i,\mathbf{r}}\left\{ [S_{i}^{x}(\mathbf{r})]{}^{2}-[S_{i}^{y}(\mathbf{r})]{}^{2}\right\} ,\label{eq:H_si}
\end{equation}
where positive values of $D$ imply easy-axis anisotropy, and negative
values of $D$ imply easy-plane anisotropy. Axial site symmetries
(hexagonal, trigonal, or tetragonal) allow only $D\neq0$, whereas
rhombic site symmetries (orthorhombic, monoclinic, and triclinic)
allow both $D\neq0$ and $E\neq0$. In Eq. (\ref{eq:H_si}), $x,$
$y$ and $z$ denote spin components with respect to mutually-orthogonal
principal axes, $\mathbf{\hat{n}}^{x},\mathbf{\hat{n}}^{y},\mathbf{\hat{n}}^{z}$,
which are ``local'' as they can be different for each magnetic site
in the unit cell. In axial site symmetries, one of the principal axes
(conventionally $\mathbf{\hat{n}}^{z}$) is aligned with the high-symmetry
local rotation axis, and the remaining two axes are mutually orthogonal
in the perpendicular plane. In orthorhombic point groups, the principal
axes are parallel to the three two-fold rotation axes. In monoclinic
point groups, one axis is parallel to the two-fold rotation axis.
Additional measurements, such as bulk susceptibility or spherical
neutron polarimetry \citep{Qureshi_2019}, are needed to indicate
an appropriate axis within the perpendicular plane. Once the local
axes have been defined for one atomic position in the unit cell, they
can be generated for all others by applying space-group symmetry operations. 

The dimensionality $n$ of the spins may be effectively reduced in
the limit of strong axial anisotropy, which occurs when $|D|\gg k_{\mathrm{B}}T$.
Then, instead of considering Heisenberg spins ($n=3$), we may consider
Ising spins ($n=1$) for strong easy-axis anisotropy, or XY spins
($n=2$) for strong easy-plane anisotropy.

\subsection{Magnetic neutron scattering intensity }

Our goal here is to express the diffuse magnetic neutron-scattering
intensity for $T>T_{\mathrm{N}}$ in terms of the underlying magnetic
interactions. This requires either analytical approximations, such
as field theories, or numerical simulations, such as Monte Carlo simulations.
Spinteract uses a self-consistent extension of mean-field theory called
Onsager reaction-field theory \citep{Brout_1967,Logan_1995,Eastwood_1995,Wysin_2000,Scherer_1976,Scherer_1977}.
This approach takes account of thermal fluctuations in an approximate
way, but has been shown to give results in excellent agreement with
Monte Carlo simulations \citep{Conlon_2010,Paddison_2020}. The equations
underlying this approach are discussed in the literature for special
cases \citep{Brout_1967,Eastwood_1995,Logan_1995,Wysin_2000,Paddison_2020};
I sketch the derivation of the general case below. The key results
are given in Eqs.~(\ref{eq:intensity_mfo})--(\ref{eq:onsager}).

I start with the general expression for the energy-integrated magnetic
neutron-scattering intensity from spin-only moments a single crystal
\citep{Lovesey_1987},

\begin{align}
I(\mathbf{Q}) & =\frac{C[gf(Q)]^{2}}{N}\sum_{i,j}\left\langle \mathbf{S}_{i}^{\perp}(-\mathbf{Q})\cdot\mathbf{S}_{j}^{\perp}(\mathbf{Q})\right\rangle \exp\left[\mathrm{i}\mathbf{Q}\cdot(\mathbf{R}_{j}-\mathbf{R}_{i})\right],\label{eq:neutron}
\end{align}
where $f(Q)=f(|\mathbf{Q}|)$ is the magnetic form factor \citep{Brown_2004},
and
\begin{equation}
\mathbf{S}_{i}^{\perp}=\mathbf{S}_{i}-\mathbf{Q}\thinspace\mathbf{S}_{i}\cdot\mathbf{Q}/Q^{2}\label{eq:s_perp}
\end{equation}
is the projection of the spin perpendicular to $\mathbf{Q}$. The
Fourier transform of a spin component is defined by
\begin{align}
S_{i}^{\alpha}(\mathbf{Q}) & =\sum_{\mathbf{r}}S_{i}^{\alpha}(\mathbf{r})\exp(\mathrm{i}\mathbf{Q}\cdot\mathbf{r})\label{eq:magn}
\end{align}
and the Fourier transform of the exchange interaction is defined by
\begin{equation}
J_{ij}^{\alpha\beta}(\mathbf{Q})=\sum_{\mathbf{r}^{\prime}-\mathbf{r}}J_{ij}^{\alpha\beta}(\mathbf{r}^{\prime}-\mathbf{r})\exp[-\mathrm{i}\mathbf{Q}\cdot(\mathbf{r}^{\prime}-\mathbf{r})],\label{eq:j_q}
\end{equation}
which is independent of the choice of the unit cell $\mathbf{r}$
at the origin. The $J_{ij}^{\alpha\beta}$ describe a $nN\times nN$
Hermitian ``interaction matrix'' at each wavevector, where $n$
is the spin dimension.

The Onsager reaction-field approach consider the spin alignment induced
by a site-dependent applied field $H_{i}^{\alpha}(\mathbf{Q})$, which
leads to an effective field at site $i$ given by
\begin{equation}
H_{i,\mathrm{eff}}^{\alpha}(\mathbf{Q})=H_{i}^{\alpha}(\mathbf{Q})-\lambda S_{i}^{\alpha}(\mathbf{Q})+\sum_{j}J_{ij}^{\alpha\beta}(\mathbf{Q})S_{j}^{\beta}(\mathbf{Q}).\label{eq:effective_field}
\end{equation}
Here, $\lambda$ is the reaction field, which is a temperature-dependent
parameter that is subtracted from the mean field to account for the
effect of local spin correlations. It is determined at each temperature
by enforcing the self-consistency condition on the average spin length,
\begin{equation}
\frac{1}{NN_{\mathbf{q}}}\sum_{\alpha,i,\mathbf{q}}\left\langle S_{i}^{\alpha}(\mathbf{q})S_{i}^{\alpha}(-\mathbf{q})\right\rangle =S(S+1),\label{eq:sum_rule}
\end{equation}
where the sum is taken over $N_{\mathbf{q}}$ wavevectors $\mathbf{q}$
in the first Brillouin zone \citep{Brout_1967,Logan_1995,Eastwood_1995,Wysin_2000}.
The mean-field approximation corresponds to $\lambda=0$ at all temperatures. 

The calculation proceeds by writing the spin in terms of normal-mode
variables that are indexed by $\mu\in\{1,nN\}$,
\begin{equation}
S_{i}^{\alpha}(\mathbf{Q})=\sum_{\mu}S_{\mu}(\mathbf{Q})U_{i\mu}^{\alpha}(\mathbf{Q}),\label{eq:modes}
\end{equation}
where $S_{\mu}$ is the amplitude of mode $\mu$. An analogous decomposition
is made for the field $H_{i}^{\alpha}(\mathbf{Q})$. The interaction
matrix is diagonalised by transforming it to normal-mode variables,
\begin{equation}
\lambda_{\mu}(\mathbf{Q})U_{i\mu}^{\alpha}(\mathbf{Q})=\sum_{j}J_{ij}^{\alpha\beta}(\mathbf{Q})U_{j\mu}^{\beta}(\mathbf{Q}),\label{eq:eigenvalue}
\end{equation}
where $\lambda_{\mu}(\mathbf{Q})$ are eigenvalues of the interaction
matrix, and the eigenvector components $U_{i\mu}^{\alpha}(\mathbf{Q})$
are normalised such that $\sum_{i,\alpha}U_{i\mu}^{\alpha}(\mathbf{Q})U_{i\nu}^{\alpha}(-\mathbf{Q})=\delta_{\mu\nu}$.
These eigenvalues and eigenvectors contain important information about
the physics of the system. In particular, for a given set of interaction
parameters, the wavevector at which the global maximum eigenvalue
$\lambda_{\mathrm{max}}$ occurs is the propagation vector of the
magnetically-ordered state that develops at $T_{N}$. In the absence
of frustration, $\lambda_{\mathrm{max}}$ occurs at a small number
of wavevectors related by symmetry. In highly-frustrated systems,
by contrast, there is a large degeneracy of wavevectors with maximum
eigenvalues close to $\lambda_{\mathrm{max}}$ \citep{Reimers_1991,Canals_2000}. 

Using the definition of the single-ion (Curie) susceptibility,
\begin{equation}
\chi_{0}=\frac{S(S+1)}{nT},\label{eq:chi_0}
\end{equation}
and Eqs.~(\ref{eq:magn})--(\ref{eq:j_q}) and (\ref{eq:modes})--(\ref{eq:chi_0})
in Eq.~(\ref{eq:effective_field}), we obtain the wavevector-dependent
magnetic susceptibility $\chi_{\mu}(\mathbf{Q})=S_{\mu}(\mathbf{Q})/H_{\mu}(\mathbf{Q})$
for each normal mode,
\begin{align}
\chi_{\mu}(\mathbf{Q}) & =\frac{\chi_{0}}{1-\chi_{0}\left[\lambda_{\mu}(\mathbf{Q})-\lambda\right]}.\label{eq:mode_suscp}
\end{align}
The magnetic ordering temperature $T_{N}$ of the model is the highest
temperature at which the denominator of Eq.~(\ref{eq:mode_suscp})
is equal to zero, for any wavevector and any mode. The wavevector-dependent
susceptibility is closely related to the correlation function of mode
amplitudes, \emph{via }the high-temperature limit of the fluctuation-dissipation
theorem \citep{Lovesey_1987},
\begin{equation}
\chi_{\mu}(\mathbf{Q})=\frac{1}{T}\left\langle S_{\mu}(\mathbf{Q})S_{\mu}(-\mathbf{Q})\right\rangle .\label{eq:fluct_diss}
\end{equation}
The final result for the diffuse scattering intensity is obtained
by using Eqs.~(\ref{eq:modes}), (\ref{eq:mode_suscp}), and (\ref{eq:fluct_diss})
in Eq.~(\ref{eq:neutron}):
\begin{equation}
I(\mathbf{Q})=\frac{C[\mu f(Q)]^{2}}{N}\sum_{\mu=1}^{nN}\frac{|\mathbf{s}_{\mu}^{\perp}(\mathbf{Q})|^{2}}{1-\chi_{0}\left[\lambda_{\mu}(\mathbf{Q})-\lambda\right]},\label{eq:intensity_mfo}
\end{equation}
in which 
\begin{equation}
\mathbf{s}_{\mu}^{\perp}(\mathbf{Q})=\sum_{i,\alpha}(\hat{\mathbf{n}}_{i}^{\alpha}-\mathbf{Q}\thinspace\hat{\mathbf{n}}_{i}^{\alpha}\cdot\mathbf{Q}/Q^{2})U_{i\mu}^{\alpha}(\mathbf{Q})\exp(\mathrm{i}\mathbf{Q}\cdot\mathbf{R}_{i})\label{eq:s_perp_mfo}
\end{equation}
and the magnetic moment $\mu=g\sqrt{S(S+1)}$. The reaction field
is determined at each temperature by requiring that
\begin{equation}
\sum_{\mu,\mathbf{q}}[1-\chi_{0}(\lambda_{\mu}(\mathbf{q)}-\lambda)]^{-1}=nNN_{\mathbf{q}},\label{eq:onsager}
\end{equation}
which follows from Eq.~(\ref{eq:sum_rule}), (\ref{eq:modes}), (\ref{eq:mode_suscp}),
and (\ref{eq:fluct_diss}).

The first term in brackets in Eq.~(\ref{eq:s_perp_mfo}) accounts
for the fact that neutrons are only sensitive to magnetisation components
perpendicular to $\mathbf{Q}$. This direction dependence allows magnetic
diffuse scattering measurements to be sensitive to the bond-dependence
of interactions, enabling investigation of non-Heisenberg interactions
\citep{Paddison_2020}. In the limit of Heisenberg spins with entirely
isotropic interactions, only one spin component needs to be considered,
and Eq.~(\ref{eq:intensity_mfo}) simplifies to

\begin{equation}
I_{\mathrm{iso}}(\mathbf{Q})=\frac{2C[\mu f(Q)]^{2}}{3N}\sum_{\mu=1}^{N}\frac{|\sum_{i}U_{i\mu}(\mathbf{Q})\exp(\mathrm{i}\mathbf{Q}\cdot\mathbf{R}_{i})|^{2}}{1-\chi_{0}\left[\lambda_{\mu}(\mathbf{Q})-\lambda\right]}.\label{eq:intensity_iso_mfo}
\end{equation}

Finally, to compare calculations with data collected on powder samples,
it is necessary to perform a spherical average of Eq.~(\ref{eq:intensity_mfo})
to obtain $I(Q)$. In Spinteract, the spherical average is performed
numerically using the method of Lebedev quadrature \citep{Lebedev_1999}.
This approach is advantageous because the distribution of samples
has $m\bar{3}m$ Laue symmetry, which allows a large reduction in
the number of wavevectors that must be sampled in high-symmetry systems.

\subsection{Magnetic susceptibility}

The bulk magnetic susceptibility $\chi$ for a powder sample is obtained
from
\begin{equation}
\chi T=\frac{\mu^{2}}{nN}\sum_{\mu=1}^{nN}\frac{|\sum_{i,\alpha}\hat{\mathbf{n}}_{i}^{\alpha}U_{i\mu}^{\alpha}(\mathbf{0})|^{2}}{1-\chi_{0}\left[\lambda_{\mu}(\mathbf{0})-\lambda\right]}\label{eq:intensity_mfo-1}
\end{equation}
in the general case, and
\begin{equation}
\chi_{\mathrm{iso}}T=\frac{\mu^{2}}{N}\sum_{\mu=1}^{N}\frac{|\sum_{i}U_{i\mu}(\mathbf{0})|^{2}}{1-\chi_{0}\left[\lambda_{\mu}(\mathbf{0})-\lambda\right]}\label{eq:suscp_iso}
\end{equation}
in the isotropic case. Since the bulk magnetic susceptibility measures
$\mathbf{Q}=\mathbf{0}$, only the normal mode corresponding to ferromagnetic
spin alignment contributes to it. The eigenvalue for this mode
\begin{align*}
\lambda_{\mathrm{bulk}} & =\sum_{n}J_{n}Z_{n}\\
 & =3\theta_{\mathrm{W}}/S(S+1),
\end{align*}
where $Z_{n}$ is the coordination number for $n$-th neighbour interactions,
and $\theta_{\mathrm{W}}$ is the Weiss temperature that is commonly
fitted to bulk magnetic susceptibility data at high temperatures \citep{Mugiraneza_2022}.
Note that Eq.~(\ref{eq:suscp_iso}) reproduces the Curie-Weiss law
in the mean-field approximation ($\lambda=0$).

\subsection{Advantages and limitations}

The reaction-field approach described has two key properties that
make it useful for fitting magnetic diffuse scattering data. First,
each scattering calculation is relatively fast, typically taking no
more than a few seconds. Second, unlike statistical approaches such
as Monte Carlo simulations, the calculation results do not contain
statistical noise. This is important because most fitting algorithms
numerically calculate derivatives of the goodness-of-fit metric as
a key step, which requires that the calculated curves are free from
noise. 

Onsager reaction-field theory, and an equivalent theory called the
self-consistent Gaussian approximation (SCGA) \citep{Conlon_2010,Plumb_2019},
have been shown to give accurate results for realistic magnetic models.
For example, a study of the frustrated Heisenberg model on the pyrochlore
lattice considered antiferromagnetic nearest-neighbour interactions
and various further neighbor interactions, and found that the scattering
patterns obtained from the SCGA were in excellent agreement with Monte
Carlo simulations over a wide temperature range \citep{Conlon_2010}.
Subsequently, reaction-field calculations involving bond-dependent
interactions on triangular and honeycomb lattices were compared with
Monte Carlo simulation results, again showing good agreement between
the two approaches \citep{Paddison_2020}. These comparisons allow
confidence in applying the approach to materials where the Hamiltonian
is not yet known.

The reaction-field approach also has some limitations. It is only
exact for classical spins in the high-temperature limit, and becomes
less accurate close to a magnetic ordering transition, although it
is notable that a reaction-field study of MnO obtained results for
$T\gtrsim T_{N}$ that were comparable to those at higher temperatures
\citep{Hohlwein_2003}. Importantly, reaction-field calculations are
much less accurate for systems with low coordination numbers, such
as quasi-one-dimensional magnets \citep{Scherer_1977,Pires_1978},
and Spinteract should be used with caution in such systems. Reaction-field
theory also does not consider effects arising from quantum fluctuations,
order-by-disorder, or partial magnetic ordering where magnetic Bragg
and diffuse scattering coexist.

\section{Implementation}

In Spinteract, interaction parameters can be fitted to multiple data
sets simultaneously, including powder and single-crystal magnetic
diffuse-scattering data, and powder bulk magnetic susceptibility.
The user provides a keyword-based text file containing crystallographic
information (unit-cell parameters and fractional coordinates of magnetic
atoms) and magnetic information (spin dimensionality $n$, spin magnitude,
magnetic form factor, and local principal axes, if required). The
user also specifies the interaction parameters included in the model
and their initial values. Spinteract can automatically implement Heisenberg
interactions $J_{\mathrm{iso}}$ for arbitrary neighbours, anisotropic
interactions $J_{xx}$, $J_{yy}$, and $J_{zz}$, the long-ranged
dipolar interaction $D_{\mathrm{dip}}$, and single-ion anisotropy
terms $D$ and $E$. If more complex interactions are required, custom
coupling matrices can be defined. Detailed instructions are provided
with Spinteract, which is open-source software and may be downloaded
from \emph{www.joepaddison.com/software}.

Spinteract comprises a Fortran subroutine that calculates the sum
of squared residuals, defined as
\begin{equation}
\chi^{2}=\sum_{\mathrm{d}}W_{\mathrm{d}}\sum_{i\in\mathrm{d}}\left(\frac{I_{i}^{\mathrm{data}}-s_{\mathrm{d}}I_{i}^{\mathrm{calc}}}{\sigma_{i}}\right)^{2},\label{eq:chisq}
\end{equation}
where $\mathrm{d}$ denotes a data set with weight $W_{\mathrm{d}}$,
$I_{i}^{\mathrm{data}}$ is the intensity of data point $i$, $I_{i}^{\mathrm{calc}}$
is the calculated intensity obtained from Eqs.~(\ref{eq:intensity_mfo})--(\ref{eq:intensity_iso_mfo}),
$\sigma_{i}$ is the corresponding uncertainty, and $s_{\mathrm{d}}$
is an (optional) refined intensity scale factor which is determined
for each dataset from the linear-least-squares relation \citep{Proffen_1997}
\begin{equation}
s_{\mathrm{d}}=\frac{\sum_{i}I_{i}^{\mathrm{data}}I_{i}^{\mathrm{calc}}/\sigma_{i}^{2}}{\sum_{i}(I_{i}^{\mathrm{calc}})^{2}/\sigma_{i}^{2}}.
\end{equation}
This is necessary because neutron-scattering data are typically not
placed in absolute intensity units of bn\,sr$^{-1}$\,spin$^{-1}$.
In Spinteract, this equation can be extended so that an intensity
offset is also refined \citep{Proffen_1997}, to account for residual
background scattering or incoherent scattering. These profile parameters
can be constrained to be the same for all datasets, if needed.

The fitting procedure---the minimisation of Eq.~(\ref{eq:chisq})
by varying the interaction parameters---is performed using the well-established
program Minuit \citep{James_1975,James_1994}. All the capabilities
of Minuit, such as estimation of parameter uncertainties and contour
plotting of $\chi^{2}$, are accessible in Spinteract. Minuit includes
several different fitting algorithms, but the most widely used is
the Migrad algorthm, which implements a variant of the Davidon-Fletcher-Powell
algorithm for non-linear least squares fitting. This algorithm was
used for the examples discussed in the following section.

A Spinteract calculation proceeds as follows First, if a centred unit
cell is specified, the unit cell is transformed to the primitive setting,
to enable calculation of the interaction matrices in their most compact
form. Second, the interaction matrix is calculated from Eq.~(\ref{eq:j_q})
for the current interaction parameters, and diagonalised on a grid
in the primitive cell in reciprocal space (equivalent to the first
Brillouin zone). Third, the reaction field is calculated at each temperature
\emph{via} Eq.~(\ref{eq:onsager}). Fourth, the scattering intensity
is calculated \emph{via} Eqs.~(\ref{eq:intensity_mfo})--(\ref{eq:intensity_iso_mfo}),
and spherically averaged if needed. Fifth, the goodness-of-fit metric
$\chi^{2}$ is obtained from Eq.~(\ref{eq:chisq}), and the optimal
intensity scale and background factors are determined. If the interactions
are being refined, $\chi^{2}$ for the current interaction parameters
is sent to the Minuit fitting program, and Minuit returns a new set
of interactions for the next iteration of the fit. 

After the fit has converged, Spinteract outputs the calculated scattering
patterns for the optimised parameter values. Spinteract also outputs
a summary of the model parameter values, uncertainties, and refinement
metrics in LaTeX format, for straightforward integration into future
publications. 

\section{Examples}

In this section, I present examples of Spinteract refinements to experimental
neutron-scattering data for two well-studied materials: the canonical
antiferromagnet MnO and the well-known frustrated magnet Gd$_{3}$Ga$_{5}$O$_{12}$.
These examples are not intended primarily to uncover new information
about these materials, but rather as test cases that demonstrate the
effectiveness of the approach. However, we will see that these refinements
allow some new information to be determined.

\subsection{Single crystal data: MnO}

\begin{figure*}
\begin{centering}
\includegraphics{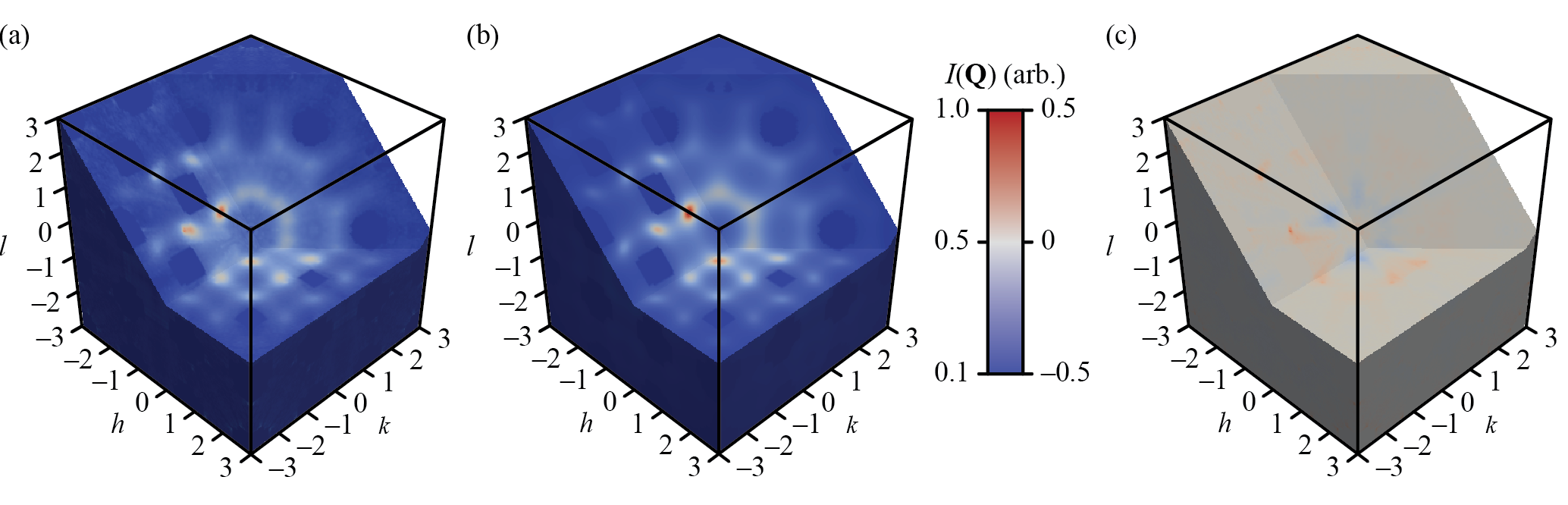}
\par\end{centering}
\centering{}\caption{\label{fig:fig1}Single-crystal diffuse magnetic scattering for MnO,
showing (a) experimental data measured at $160$\,K (reproduced from
Ref.~\citep{Paddison_2018}); (b) fit obtained using Spinteract;
(c) data--fit.}
\end{figure*}

Manganese(II) oxide, MnO, plays a central role in the history of neutron
scattering---it was the first material in which antiferromagnetic
Bragg peaks were observed below its $T_{N}$ of $118$\,K, providing
the first direct experimental evidence for antiferromagnetism \citep{Shull_1949}.
In the 1960s, magnetic diffuse scattering was reported above $T_{N}$.
An early powder diffuse-scattering analysis identified that the dominant
magnetic interactions between $S=5/2$ magnetic moments of Mn$^{2+}$
was the next-nearest neighbour Heisenberg term, which was estimated
to have an antiferromagnetic value of $J_{2}=4.65$\,meV \citep{Blech_1964}.
Subsequently, an analysis of single-crystal diffuse scattering obtained
antiferromagnetic $J_{1}=3.31$\,K and $J_{2}=4.69$\,K at $T\approx160$\,K
using the Onsager reaction-field approach \citep{Hohlwein_2003},
which showed reasonable consistency with values extrapolated from
low-temperature spin-wave-dispersion measurements \citep{Pepy_1974}.
MnO has also been investigated by reverse Monte Carlo analysis of
powder \citep{Mellergard_1998} and single-crystal \citep{Paddison_2018}
samples, as well as by magnetic pair-distribution-function analysis
of powder samples \citep{Frandsen_2015}, which confirm that antiferromagnetic
correlations and domain structure persist to temperatures far above
$T_{N}$.

Using the Spinteract program, refinements of $J_{1}$ and $J_{2}$
were performed against experimental magnetic diffuse scattering data
on MnO. The data were previously published in Ref.~\citep{Paddison_2018};
they were collected at $T=160$\,K using the SXD diffractometer at
the ISIS Neutron and Muon Source, and comprise a large volume of reciprocal
space ($\sim3\times10^{6}$ data points), as shown in Fig.~\ref{fig:fig1}.
Nuclear Bragg peaks were excluded and a measurement of an empty sample
holder was subtracted from the data. Spinteract refinements of $J_{1}$
and $J_{2}$ to the entire volume of data were performed. An overall
intensity scale factor and a constant-in-$Q$ offset were also refined;
the latter was needed to account for significant incoherent scattering
from Mn. Despite the large number of data points, convergence was
obtained after a CPU time of approximately $10$ minutes on a 2.9
GHz Intel Core i5 processor. As shown in Fig.~\ref{fig:fig1}, an
excellent fit was obtained with the refined values of $J_{1}$ and
$J_{2}$ given in Table~\ref{tab:MnO_params}; these values show
excellent agreement (within a few per cent) with the results of Ref.~\citep{Hohlwein_2003}.
Since statistical uncertainties were not reported in the original
data set, parameter uncertainties could not readily be estimated.
However, a further refinement in which a linear-in-$Q$ intensity
offset was also refined yielded essentially equivalent parameter values
{[}Table~\ref{tab:MnO_params}{]}, suggesting that the results are
likely to be robust.

A natural question is whether the magnetic interactions in MnO might
extend beyond next-nearest-neighbours. To investigate this question,
a further refinement was performed, in which $J_{3}$ was allowed
to vary in addition to $J_{1}$ and $J_{2}$. The refined value of
$J_{3}$ is only 2\% of $J_{2}$ {[}Table~\ref{tab:MnO_params}{]},
suggesting that interactions beyond next-nearest neighbours are indeed
very weak in MnO.

\begin{table}
\begin{centering}
\begin{tabular}{c|cccc}
\hline 
MnO & $J_{1}$ (K) & $J_{2}$ (K) & $J_{3}$ (K) & $R_{\mathrm{wp}}$\tabularnewline
\hline 
\hline 
$160$\,K (constant offset) & $3.26$ & $4.45$ & $0^{\ast}$ & $8.76$\tabularnewline
$160$\,K (constant offset) & $3.44$ & $4.48$ & $0.09$ & $8.76$\tabularnewline
$160$\,K (constant + $Q$-linear offset) & $3.33$ & $4.51$ & $0^{\ast}$ & $8.75$\tabularnewline
$160\pm20$\,K (Ref.~\citep{Hohlwein_2003}) & $3.31$ & $4.59$ & $0^{\ast}$ & $-$\tabularnewline
\hline 
\end{tabular}
\par\end{centering}
\caption{\label{tab:MnO_params}Values of interaction parameters for MnO, obtained
from refinements to single-crystal diffuse scattering data shown in
Figure~\ref{fig:fig1}, compared with values obtained in Ref.~\citep{Hohlwein_2003}.
Interaction parameters are given for spins of magnitude $\sqrt{S(S+1)}$
with $S=5/2$, and in the same Hamiltonian convention as previous
studies, where antiferromagnetic interactions are positive and spin
pairs are double counted. In the Spinteract convention defined by
Eq.~(\ref{eq:heisenberg}), the interaction parameters in this table
would be multiplied by $-2$.}
\end{table}

\subsection{Powder data: Gd$_{3}$Ga$_{5}$O$_{12}$}

\begin{figure*}
\includegraphics{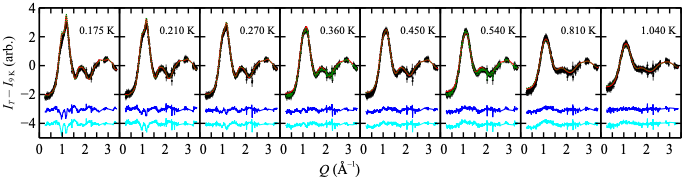}
\centering{}\caption{\label{fig:fig2}Powder diffuse magnetic scattering for Gd$_{3}$Ga$_{5}$O$_{12}$
at temperatures labelled in each panel. Experimental data (black circles)
are reproduced from Refs.~\citep{Petrenko_1998,Paddison_2015}. Spinteract
fits with variable $J_{1},$ $J_{2}$, $J_{3}$, and $J_{4}$ are
shown as red lines, and the corresponding difference curves (data--fit)
are shown as blue lines that are vertically shifted by 3 units. Spinteract
fits with variable $J_{2}$ and $J_{3}$ and fixed $J_{1}=0.107$\ K
and $J_{4}=0$ are shown as dotted green lines, and the corresponding
difference curves (data--fit) are shown as cyan lines that are vertically
shifted by 4 units. }
\end{figure*}

Garnet-structured Gd$_{3}$Ga$_{5}$O$_{12}$ (gadolinium gallium
garnet, GGG) is a geometrically-frustrated system in which Gd$^{3+}$
moments with $S=7/2$ occupy two interpenetrating networks of corner-sharing
triangles connected in three dimensions. It shows an unusual spin-freezing
transition below $T_{f}\approx0.14$\,K \citep{Schiffer_1995}, and
the emergence of effective magnetic multipoles from strongly-correlated
10-spin loops above $T_{f}$ \citep{Paddison_2015}. Its magnetic
interactions consist of the long-ranged magnetic dipolar interaction,
with magnitude $D_{\mathrm{dip}}=0.0457$\,K at the nearest-neighbour
distance, and local exchange interactions extending at least to third-nearest
neighbours \citep{Kinney_1979,Yavorskii_2006,dAmbrumenil_2015}. The
values of the exchange interactions were first investigated using
a high-temperature analysis, which obtained antiferromagnetic $J_{1}=0.107$\,K,
$-0.015\leq J_{2}\leq0.009$\,K, and $-0.03\leq J_{3}\leq0.100$\,K
\citep{Kinney_1979}. A later study of magnetic critical scattering
considered the incipient ordering wavevector within the spin-frozen
regime, and obtained $-0.012\leq J_{2}\leq-0.004$\,K and $-0.003\leq J_{3}\leq0.012$\,K
\citep{Yavorskii_2006}. A spin-wave measurement in applied magnetic
field was consistent with $J_{3}\approx0.013$\,K and insensitive
to $J_{2}$ \citep{dAmbrumenil_2015}. 

Spinteract refinements were performed to published powder neutron-scattering
data collected over a wide temperature range above $T_{f}$, as shown
in Fig.~\ref{fig:fig2}. These data were originally published in
Ref.~\citep{Petrenko_1998}. Initially, $J_{1}$ was kept fixed at
the value of $0.107$\,K \citep{Kinney_1979,Yavorskii_2006} and
$J_{2}$ and $J_{3}$ were refined, as in Ref.~\citep{Yavorskii_2006}.
The refined fit parameters are given in Table~\ref{tab:ggg_params};
they are close to the possible range reported in Ref.~\citep{Yavorskii_2006},
with $J_{2}$ at the upper limit of this range and $J_{3}$ close
to its lower limit. However, within this range, optimal values of
$J_{2}\approx-0.005$\,K and $J_{3}\approx0.010$\,K were also reported
in Ref.~\citep{Yavorskii_2006}; the latter value is significantly
different to the value from Spinteract refinement. This difference
is mainly because reaction-field theory predicts an onset of long-range
magnetic ordering at $T>0.2$\,K for the optimal parameter set of
Ref.~\citep{Yavorskii_2006}, which is inconsistent with the observation
of diffuse magnetic scattering below this temperature. An extended
model was therefore tested using Spinteract, where $J_{1}$, $J_{2}$,
$J_{3}$ and $J_{4}$ were allowed to vary. This refinement yielded
a significant improvement in fit quality {[}Fig.~\ref{fig:fig2}{]}
and an antiferromagnetic value of $J_{3}$. However, the refined value
of $J_{4}$ is also significant, suggesting the possibility that interactions
may extend beyond third-nearest neighbours in Gd$_{3}$Ga$_{5}$O$_{12}$.

\begin{table}
\begin{centering}
\begin{tabular}{c|ccccc}
\hline 
{\footnotesize{}Gd$_{3}$Ga$_{5}$O$_{12}$ } & {\footnotesize{}$J_{1}$ (K)} & {\footnotesize{}$J_{2}$ (K)} & {\footnotesize{}$J_{3}$ (K)} & {\footnotesize{}$J_{4}$ (K)} & {\footnotesize{}$R_{\mathrm{wp}}$}\tabularnewline
\hline 
{\footnotesize{}Fit $J_{2}$, $J_{3}$} & {\footnotesize{}$0.107^{\ast}$} & {\footnotesize{}$-0.0035(2)$} & {\footnotesize{}$-0.0015(7)$} & {\footnotesize{}$0^{*}$} & {\footnotesize{}$13.28$}\tabularnewline
{\footnotesize{}Fit $J_{1}$--$J_{4}$} & {\footnotesize{}$0.130(2)$} & {\footnotesize{}$-0.0038(1)$} & {\footnotesize{}$0.0040(6)$} & {\footnotesize{}$0.0040(3)$} & {\footnotesize{}$12.62$}\tabularnewline
{\footnotesize{}Ref.~\citep{Yavorskii_2006}} & {\footnotesize{}$0.107^{\ast}$} & {\footnotesize{}$-0.012:-0.004$} & {\footnotesize{}$-0.003:0.012$ } & {\footnotesize{}$0^{*}$} & {\footnotesize{}$-$}\tabularnewline
\hline 
\end{tabular}
\par\end{centering}
\caption{\label{tab:ggg_params}Values of interaction parameters for Gd$_{3}$Ga$_{5}$O$_{12}$,
obtained from refinements to powder diffuse-scattering data shown
in Figure~\ref{fig:fig1}, compared with range of possible values
obtained in Ref.~\citep{Yavorskii_2006}. In all cases, $D_{\mathrm{dip}}=0.0457$\,K
is fixed. Interaction parameters are given for spins of magnitude
$\sqrt{S(S+1)}$ with $S=7/2$, and in the same Hamiltonian convention
as previous studies \citep{Yavorskii_2006,Kinney_1979}, where antiferromagnetic
interactions are positive and spin pairs are single counted. For the
Spinteract convention defined by Eq.~(\ref{eq:heisenberg}), the
interaction parameters in this table would be multiplied by $-1$.}
\end{table}

\section{Guidance}

The Spinteract code has already been used in several published studies
\citep{Pokharel_2020,Paddison_2020,Paddison_2022,Welch_2022}, allowing
it to be tested on a variety of problems. Some general strategies
for data collection, processing, and refinement are summarised below.

\subsection{Data collection}

A key advantage of magnetic diffuse scattering is that it can often
be collected using the same experimental setup as for conventional
Bragg diffraction measurements, by increasing the sample temperature
above $T_{N}$. Standard data corrections are required, such as for
detector efficiency and absorption by the sample. For quantitative
analysis of magnetic diffuse scattering data, several other considerations
can be relevant:
\begin{itemize}
\item \emph{Sample size, counting time, and choice of instrument. }Since
diffuse scattering intensity is distributed throughout reciprocal
space, the scattering intensity at any given position is typically
weak. Hence, larger sample sizes and counting times may be required
compared to conventional Bragg diffraction measurements. For single-crystal
measurements, instruments capable of measuring a wide range of reciprocal
space are usually most suitable. 
\item \emph{Energy integration. }Spinteract assumes that the neutron-diffraction
data are energy-integrated over the entire spin-fluctuation spectrum.
This requirement is met in diffraction experiments provided that the
energy change of the scattered neutrons is much smaller than the incident
neutron energy, $E_{i}$ (``quasistatic approximation''). As a rule
of thumb, one should choose $E_{i}>k_{\mathrm{B}}\theta_{\mathrm{W}}$,
where the Weiss temperature $\theta_{\mathrm{W}}$ provides an estimate
of the strength of the magnetic interactions. 
\item \emph{Choice of measurement temperatures. }Measuring the diffuse scattering
at several temperatures above $T_{N}$ provides is often helpful to
determine the values of magnetic interactions, especially for complex
model Hamiltonians such as those considered in Refs.~\citep{Pokharel_2020,Paddison_2020,Paddison_2022,Welch_2022}.
\end{itemize}

\subsection{Data processing}

Spinteract assumes that the input data contains \emph{only} magnetic
diffuse scattering. Therefore, other signals---such as nuclear Bragg
scattering, incoherent scattering, and background scattering---should
be removed before attempting a refinement.
\begin{itemize}
\item \emph{Background subtraction. }It is important to correct diffuse-scattering
data for background scattering, which could otherwise bias the fit
results. This can be achieved by subtracting a measurement of the
empty sample container from the sample measurement. Alternatively,
the sample may be measured at a high temperature at which the spins
are essentially uncorrelated, and this measurement subtracted from
the data of interest. Since temperature-subtracted data reflect only
a difference in spin correlation between the higher and lower temperature,
this approach involves some information loss, but it is often effective
in practice.
\item \emph{Removal of nonmagnetic scattering. }It is necessary to remove
nonmagnetic scattering (e.g., nuclear Bragg peaks) from the data before
modelling them using Spinteract. This can be achieved experimentally
by performing polarisation analysis. Alternatively, unpolarised data
may be post-processed, either by subtracting a high-temperature data
set from the data of interest, by excluding regions of the data containing
Bragg peaks, or by performing a refinement to the nuclear Bragg profile
and subtracting the fitted Bragg profile from the experimental data.
\end{itemize}

\subsection{Data refinement}

The fitting program (Minuit) used in Spinteract is highly robust;
however, like any such algorithm, it can converge to a local minimum
in the goodness-of-fit, or fail to converge at all. This problem is
more likely to occur if the initial parameter values are too far from
the optimal ones, or if the parameter values are under-constrained
by the data. Some straightforward checks can help determine if the
optimal solution was found:
\begin{itemize}
\item \emph{Multiple refinements. }It is useful to perform multiple (e.g.,
10 to 100) refinements, with different starting values of the interaction
parameters. This approach often allows false (local) minima to be
identified; such false minima can be neglected provided they yield
much worse fits than the optimal solution. 
\item \emph{Covariance matrix. }A second useful check is to examine the
covariance matrix that is output by Minuit. If two (or more) parameters
have a large covariance, their values may not be well determined,
and it is worthwhile to check the fit dependence on these parameters
using a contour plot of $\chi^{2}$.
\item \emph{Physical predictions. }Spinteract provides an estimate of $T_{N}$
of the system, and the magnetic propagation vector of the ordered
state that develops below $T_{N}$, based on the refined interaction
model. For materials that exhibit long-range magnetic ordering, these
predictions can be compared with experimental results, providing an
independent check on the validity of the refined model.
\item A final check on the model validity is to repeat the refinement under
slightly different conditions---e.g., changing the relative weighting
of susceptibility \emph{vs.} diffuse-scattering data, or by allowing
an extra background parameter to refine. This can provide some insight
into the variation of parameter values due to systematic errors, which
are often more difficult to characterise than statistical uncertainties. 
\end{itemize}
If these checks suggest that the results are not well constrained
by the data, the user has two options---either fix some parameters,
or include more experimental data. A particularly useful approach
is to fit to bulk magnetic susceptibility data as well as magnetic
diffuse scattering, as in Refs.~\citep{Bai_2019,Paddison_2022,Welch_2022}.
The magnetic susceptibility expressed as $\chi T$ effectively measures
the $Q\rightarrow0$ limit of the magnetic diffuse scattering, and
provides a strong constraint on the net value of the magnetic interactions
(i.e., the Weiss temperature). Including the susceptibility data with
a sufficiently high weight can therefore help the refinement avoid
unphysical regions of parameter space.

\section{Conclusions and Outlook}

The key result of this work is to provide a user-friendly computer
program, Spinteract, for refinement of magnetic interactions against
magnetic-diffuse scattering data. The Spinteract program allows straightforward
estimation of interaction parameters from data collected in neutron-diffraction
experiments. This approach has potential to accelerate the experimental
determination of magnetic interaction in new materials, particularly
those which cannot be prepared as large single crystals suitable for
inelastic scattering measurements. Moreover, diffuse-scattering analysis
can offer advantages over conventional spin-wave analysis in topical
systems, such as those with spin-liquid or spin-glass ground states,
or materials where a strong Ising-like magnetic anisotropy generates
essentially non-dispersive spin-wave excitations \citep{Ikeda_1978}. 

The capabilities of Spinteract could be developed in several directions.
First, the equations outlined in Section~\ref{sec:Theory} assume
that magnetic spins transform as dipoles. This is the most common
case, and leads to a Curie-law form of the single-ion susceptibility.
However, spin-orbit coupling and crystal-field effects may lead to
interactions between multipolar degrees of freedom \citep{Dahlbom_2022}.
A simple example is a system in which the crystal field ground state
is two singlets, separated by an energy gap $\Delta$, in which dipolar
order only develops if $J\gtrsim\Delta$ and the transverse pseudo-spin
components transform as quadrupoles \citep{Wang_1968}. Such materials
can be modelled within the reaction-field framework by modifying the
single-ion susceptibility \citep{Santos_1980}. 

A second outstanding question is the extent to which this approach
is applicable for correlated quantum systems, such as quantum spin-liquid
candidates with effective spin-$\frac{1}{2}$. Even though the Onsager
reaction-field method is semiclassical, it does not necessarily fail
to describe the magnetic diffuse scattering of such systems, because
thermal fluctuations dominate quantum fluctuations at sufficiently
high temperature. In this way, refinements to diffuse-scattering data
at high temperature may provide information about the interactions
responsible for driving the system to a quantum ground state. Recently,
this approach was tested in the delafossite system KYbSe$_{2}$, where
Yb$^{3+}$ magnetic moments with effective spin-$\frac{1}{2}$ occupy
a triangular lattice \citep{Scheie_2021,Scheie_2022}. Refinements
to magnetic diffuse-scattering data at $T\geq1$ K using the reaction-field
approach yielded a ratio $J_{1}/J_{2}$ that is in good agreement
with advanced quantum calculations \citep{Scheie_2022}, suggesting
this approach deserves further investigation.

\section*{Acknowledgements}

I am grateful to Oleg Petrenko (Warwick) and Matthias Gutmann (ISIS)
for allowing re-use of their published diffuse-scattering data, and
to Xiaojian Bai (Louisiana State), Stuart Calder (ORNL), Andrew Christianson
(ORNL), Matthew Cliffe (Nottingham), Ovidiu Garlea (ORNL), Andrew
Goodwin (Oxford), Martin Mourigal (Georgia Tech), Ross Stewart (ISIS),
and Matthew Tucker (ORNL) for valuable discussions. Development of
the Spinteract program was supported by a Junior Research Fellowship
from Churchill College, University of Cambridge (U.K.) from 2016--2019.
Manuscript preparation was supported by the U.S. Department of Energy,
Office of Science, Basic Energy Sciences, Materials Sciences and Engineering
Division.


\end{document}